\documentclass[prb,showpacs,preprint,amsmath,amssymb,superscriptaddress]{revtex4}
\usepackage{epsfig}
\usepackage{graphicx}
\usepackage{dcolumn}
\usepackage{bm}

\begin{document}
\title{Electronic structure, magnetic and optical properties of intermetallic compounds 
R$_2$Fe$_{17}$ (R~=~Pr,~Gd)}
\author{Yu.V.~Knyazev}
\affiliation{Institute of Metal Physics, Russian Academy of Sciences-Ural Division,
620041 Yekaterinburg GSP-170, Russia}
\author{A.V.~Lukoyanov}
\affiliation{Institute of Metal Physics, Russian Academy of Sciences-Ural Division,
620041 Yekaterinburg GSP-170, Russia}
\affiliation{Ural State Technical University-UPI,
620002 Yekaterinburg, Russia}
\author{Yu.I.~Kuz'min}
\affiliation{Institute of Metal Physics, Russian Academy of Sciences-Ural Division,
620041 Yekaterinburg GSP-170, Russia}
\author{A.G.~Kuchin}
\affiliation{Institute of Metal Physics, Russian Academy of
Sciences-Ural Division, 620041 Yekaterinburg GSP-170, Russia}
\author{I.A.~Nekrasov}
\affiliation{Institute of Metal Physics, Russian Academy of
Sciences-Ural Division, 620041 Yekaterinburg GSP-170, Russia}
\affiliation{Institute of Electrophysics, Russian Academy of
Sciences-Ural Division, 620046 Yekaterinburg, Russia}

\date{\today}

\begin{abstract} 

In this paper we report comprehensive experimental and theoretical investigation of
magnetic and electronic properties of the intermetallic compounds Pr$_2$Fe$_{17}$ and Gd$_2$Fe$_{17}$.
For the first time electronic structure of these two systems was probed by optical
measurements in the spectral range of 0.22--15 $\mu$m. On top of that charge carriers
parameters (plasma frequency $\Omega$ and relaxation frequency $\gamma$) and optical
conductivity $\sigma (\omega)$ were determined. Self-consistent spin-resolved
bandstructure calculations within the conventional LSDA+U method were performed. Theoretical
interpetation of the experimental $\sigma (\omega)$ dispersions indicates transitions
between $3d$ and $4p$ states of Fe ions to be the biggest ones.
Qualitatively the line shape of the theoretical optical conductivity
coincides well with our experimental data.
Calculated by LSDA+U method magnetic moments per formula unit are found to
be in good agreement with observed experimental values of saturation magnetization.

\end{abstract}


\maketitle 
\section{Introduction}

During last two decades Fe-rich intermetallic systems stays under investigation from both
experimental and theoretical points of view because its anomalous magnetic properties. At
present the intermetallic compounds with rare-earth elements draw a significant interest
because of their well pronounced Invar properties and also as possible new cheap
high-energy storing materials for permanent 
magnets.~\cite{Coey,Mooij,Gubbens,Jacobs,Wang,Shen,Kuchin01} Rare-earth intermetallic
compounds R$_2$Fe$_{17}$ (R -- a rare-earth ion) are well-known due to large magnetic
moment of Fe ions and yet their comparatively low Curie temperature $T_C$. Further
increase of the $T_C$ of the compounds can be achieved by implantation of interstitial
atoms~\cite{Coey,Mooij} or introducing a low content of non-magnetic impurities
substituting Fe atoms.\cite{Gubbens,Jacobs,Wang,Shen,Kuchin01} In such extended systems,
compared to the parent R$_2$Fe$_{17}$ compounds, upon the doping $T_C$ grows more than
twice in some cases. This increment of $T_C$ was explained primarily as a result of the
lattice expansion with substitution of Fe ions for larger radii ions. According to this
concept, Fe-Fe interactions are ferromagnetic or antiferromagnetic for interatomic
distance larger or smaller then the critical value 0.25 nm.\cite{Givord} However, a
number of experiments~\cite{van Mens,Li,Kuchin02} showed that in the substitutional
R$_2$Fe$_{17-x}$Si$_x$ alloys the crystal lattice contracts and the magnetic moment per
unit volume decreases, while $T_C$ grows. Later experimental facts are in apparent
contradiction with the model.~\cite{Givord} Clearly, a simple approach based on
distance-dependent exchange interaction is not sufficient to explain the changes of
magnetic properties of these compounds with substitution of Fe for non-magnetic ions. 

Past life several results of band structure calculations of the systems from the family
are available.
For instance the spin-polarized calculations of electron spectra of
Y$_2$Fe$_{17}$N$_3$,\cite{Jaswal} Sm$_2$Fe$_{17-x}$M$_x$ (M=Al, Ga, Si),\cite{Sabirianov}
and Nd$_2$Fe$_{17-x}$M$_x$ (M = Si, Ga)~\cite{Huang} compounds showed that non-magnetic
impurities modify the shape and width of the spin-up and spin-down densities of states
$N(E)$. Jaswal $et$~$al$.~\cite{Jaswal} suggested to explain the increase of $T_C$ in
Y$_2$Fe$_{17}$N$_3$ via changes of the density of states at the Fermi level $N(E_F)$
owing to the lattice expansion by means of the Mohn-Wohlfarth static spin-fluctuation
model.~\cite{Mohn} An analysis of the experimental
optical,~\cite{Knyazev01,Knyazev02,Kuchin01} low-temperature heat capacity\cite{Kuchin01}
and photoemission\cite{Woods} data of some pseudobinary alloys R$_2$Fe$_{17-x}$M$_x$
with M = Al, Si revealed a qualitative correlation between the Curie temperature and 
parameters of electronic structure in frame of such an approach. Recent
experimentally measured optical conductivity of Ce$_2$Fe$_{17}$ was interpreted in terms of
band structure obtained within local density approximation (LDA).\cite{Nekrasov04}

In accordance with the foregoing systematic study of the electronic structure of
R$_2$Fe$_{17}$ compounds and their modifications with substituted Fe ions
further investigations are of fundamental importance. In this work theoretical calculations of the
electronic properties together with experimental magnetic and optical measurements
were performed.  The paper is organized as follows: in the section~\ref{exp} experimental
details and results for Pr$_2$Fe$_{17}$ and Gd$_2$Fe$_{17}$ are presented. For instance,
subsection~\ref{mag_mes} is devoted to sample preparation, X-ray diffraction structural
analysis and magnetic measurements conditions. Subsection~\ref{opt_mes} provides
description of optical experiments. Section~\ref{LSDAU} contains results of
LSDA+U~\cite{Anisimov} computations of electronic structure and magnetic properties of Pr-Fe
and Gd-Fe systems. Structure of experimentally observed optical conductivity curves is
anatomized in the section~\ref{opt_theor}. At the end we briefly summarize our paper with
the section~\ref{summary}.

\section{Experiment}
\label{exp}

\subsection{Samples and magnetic measurements}
\label{mag_mes}

The compounds Pr$_2$Fe$_{17}$ and Gd$_2$Fe$_{17}$ were prepared by induction melting in
alumina crucible under argon atmosphere. The ingots were homogenized in the high-purity
argon atmosphere at $\sim$1300~K. The purity of the alloys was checked using standard
X-ray diffractometry in Cu K$\alpha$ radiation. The samples were found to be
polycrystalline, single-phase and have rhombohedral structure of the
$Th_2Zn_{17}$-type\cite{cr_str1} (space group R$\bar{3}$m) for Pr$_2$Fe$_{17}$ and
hexagonal crystal structure of the $Th_2Ni_{17}$-type\cite{cr_str2} (space group
P6$_3$/mmc) for Gd$_2$Fe$_{17}$. The measured lattice parameters $a$ and $c$ are close to
those published earlier\cite{cr_str1,cr_str2} and are shown in the Table~\ref{tab1}.
Spherical specimens of 2-3~mm in diameter were used for magnetic measurements. For
following optical studies the specular surface of the samples was prepared by mechanical
polishing with diamond pastes.

The Curie points of the compounds were determined from temperature dependencies of $ac$
susceptibility. The $ac$ susceptibility was measured by a differential method in an $ac$
magnetic field of 8~Oe with a frequency of 80~Hz. The saturation magnetizations at 
$T$=4.2~K were determined from the isothermal magnetization measurements carried out by
means of vibrating sample magnetometer in magnetic fields up to 20 kOe. The magnetic
parameters of the compounds are listed in Table~\ref{tab1}. They are a little different
from previously reported data.~\cite{recent1,recent2} The differences among authors may
originate from some deviations of their samples from stoichiometry. 

\begin{table*}
\caption{Experimental structural, magnetic, and electronic parameters of the
intermetallic compounds Pr$_2$Fe$_{17}$ and Gd$_2$Fe$_{17}$:
lattice parameters $a$, $c$; unit-cell volume at room temperature $V$; 
$T_C$ -- Curie temperature; $M_s$ -- spontaneous magnetization at $T$ = 4.2 K;
relaxation $\gamma$ and plasma $\Omega$ frequencies; 
$N_{eff}$ -- effective concentration of conduction 
electrons.}
\begin{ruledtabular}
\begin{tabular}{ccrcccccc}
Compound & $a$ ($\AA$)&$c$ ($\AA$)&$V$ ($\AA^3$)&$T_C$~(K)&$M_s$ ($\mu_B$/f.u.)&$\gamma$ (10$^{13}$s$^{-1}$) 
& $\Omega^2$ (10$^{30}$s$^{-2}$) & $N_{eff}$ (10$^{22}$ cm$^{-3}$)\\
\hline
Pr$_2$Fe$_{17}$ & 8.579 & 12.472 & 795.0 & 294 & 36.1 & 1.9 & 21.3 & 0.67\\
Gd$_2$Fe$_{17}$ & 8.496 & 8.341 & 521.4 & 466 & 21.2 & 1.5 & 19.5 & 0.61\\
\end{tabular}
\end{ruledtabular}
\label{tab1}
\end{table*}

\subsection{Optical measurements}
\label{opt_mes}

Investigation of optical properties of Pr$_2$Fe$_{17}$ and Gd$_2$Fe$_{17}$ were carried
out at room temperature by ellipsometric Beattie technique.~\cite{Beattie} Spectroscopic
ellipsometry is based on the fact that the state of polarization of incident light is
changed on reflection. The optical constants -- refractive index $n$ and absorption
coefficient $k$ -- were measured in the spectral range of $\hbar\omega$=0.077--5.6~eV ($\omega$
is a cyclic frequency of light) with accuracy of 2--4\%. From $n$ and $k$, the real
$\epsilon_1(\omega)=n^2-k^2$ and imaginary $\epsilon_2(\omega)=2nk$ parts of the
complex dielectric function $\epsilon(\omega)$, the optical conductivity 
$\sigma(\omega)=nk\omega/2\pi$, and the reflectance 
$R(\omega)=[(n-1)^2+k^2]/[(n+1)^2+k^2]$ were derived. Measurements of reflection spectra
followed by the Kramers-Kronig analysis were applied to determine the optical parameters
in the short-wave range ($\hbar\omega$=5.6--8.5~eV).

\begin{figure}[]
\begin{center}
\epsfxsize=8.6cm
\epsfbox{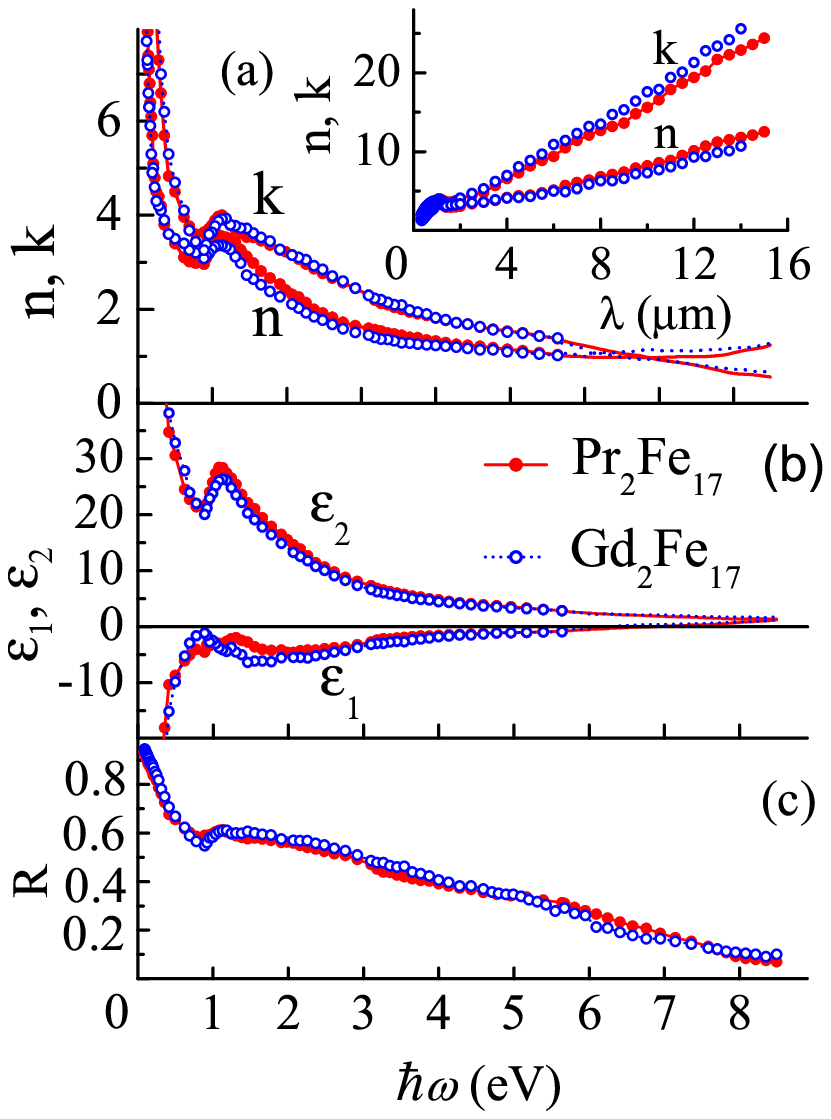}
\end{center}
\caption{(Colour online) Experimental optical constants $n$ and $k$ (panel a), dielectric
functions  $\epsilon_1$ and $\epsilon_2$ (panel b), and reflectivity spectra $R$ (panel
c) for Pr$_2$Fe$_{17}$ (full circles and solid line) and Gd$_2$Fe$_{17}$ (empty circles
and doted line) compounds. Doted and solid lines above 5.6 eV represent values obtained
by Kramers-Kronig transformation from reflection spectra. Inset presents dependences of
the optical constants on wavelength for an expanded view in infrared region.}
\label{fig1}
\end{figure}

The results of optical study ($n, k, \epsilon_1, \epsilon_2, R$ as functions of $\omega$)
for the Pr and Gd compounds are shown in Fig.~\ref{fig1}. It is seen that the optical
properties of both compounds are rather similar. All directly measured dispersions and
further derived quantities are characterized by the broad feature with maximum at the
photon energies near 1~eV. As it follows from the lineshape of the $\epsilon_2(\omega)$,
there is a strong absorption region at energies $\hbar\omega>$1~eV. With increase of the
wavelength to infrared range ($\lambda\geq$2~$\mu$m) (inset of Fig.~\ref{fig1}a) the
nonmonotonic behaviour of $n$ and $k$ is changed by a smooth growth related to the
domination of free-electron absorption. Such a character of the frequency dependencies
for the optical constants together with negative quantities $\epsilon_1$, as a rule, is
typical for the metal-like solids. As one can see from dispersion curves $R(\omega)$, in
the low-energy range the reflectance exhibits the high values. The analysis of the energy
dependence of $\epsilon_1$ and $\epsilon_2$ in this spectral interval, corresponding to
intraband electronic excitations, makes it possible to determine the plasma frequency
$\Omega$ and the effective relaxation frequency $\gamma$ of free charge carriers. Within
the assumption that light absorption for these energies has Drude character, the
parameters $\Omega^2=\omega^2(\epsilon_1^2+\epsilon_2^2)/\epsilon_1$ and 
$\gamma=\epsilon_2\omega/\epsilon_1$ were computed. In the long-wave region 
$\lambda>$8~$\mu$m, $\gamma$ and $\Omega^2$ become frequency independent. The values of
$\gamma$ and $\Omega^2$ in this energy interval were used then to estimate the effective
concentration of conduction electrons $N_{eff}=\Omega^2m/4\pi e^2$ ($m$ and $e$ are the
mass and the charge of a free electron respectively). All these parameters obtained from
experimental data treatment are presented in Table~\ref{tab1}.

\begin{figure*}[]
\begin{center}
\epsfxsize=17.8cm
\epsfbox{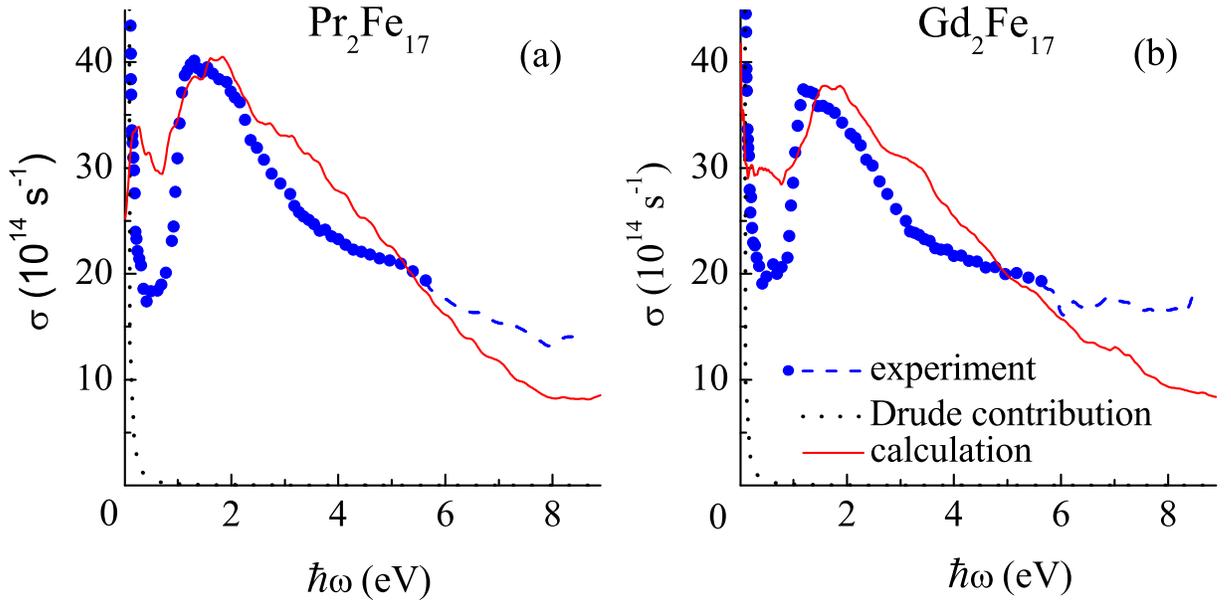}
\end{center}
\caption{(Colour online) Experimental (dark circles and dashed lines) and calculated
(solid lines) dispersion dependencies of optical conductivity for Pr$_2$Fe$_{17}$ (a) and
Gd$_2$Fe$_{17}$ (b) compounds. Dashed lines represent $\sigma(\omega)$ values obtained by
Kramers-Kronig transformation from reflection spectra. 
Doted lines correspond to intraband contributions estimated by Drude formula.
Calculated values are in arbitrary units.}
\label{fig2}
\end{figure*}

Optical conductivities $\sigma(\omega)$ for both compounds are displayed in
Fig.~\ref{fig2}. A monotonic increment of the experimental $\sigma(\omega)$ dispersion
observed in the low-energy range is related to the Drude mechanism
of electron excitations. 
For both systems intraband (Drude-like) contribution to the optical conductivity were computed according to the relation $\sigma_{Intra}(\omega)=\Omega^2\gamma/4\pi(\omega^2+\gamma^2)$ and drown in Fig.~\ref{fig2} black dotted lines. 
Corresponding values of $\Omega^2$ and $\gamma$ are given in Table~\ref{tab1}.
The magnitude of these contributions falls down sharply with energy and becomes insignificant 
above 0.5 eV. The values of static conductivity $\sigma_{Intra}(0)$ are 
estimated to be 0.89*10$^{16}$ s$^{-1}$ for Pr$_2$Fe$_{17}$ and 1.03*10$^{16}$ s$^{-1}$ for 
Gd$_2$Fe$_{17}$ correspondingly.
With increase of photon energy the $\sigma(\omega)$ curves show
a very intense asymmetric stuctures in the near infrared region of spectra at
$\sim$1.2~eV. These structures have a pronounced shoulder on high-energy side and abrupt
low-energy edge. Such a behaviour is a typical manifestation of the predominance of the
interband absorption in this energy interval. The similar shape of the $\sigma(\omega)$
curves was early observed in Y$_2$Fe$_{17}$, Ce$_2$Fe$_{17}$, and Lu$_2$Fe$_{17}$
compounds.~\cite{Knyazev02}

\section{Electronic structure calculations}
\label{LSDAU}

To calculate electronic structure of the intermetallic compounds under investigation
LSDA+U method~\cite{Anisimov} within the TB-LMTO-ASA package (Tight Binding, Linear
Muffin-Tin Orbitals, Atomic Sphere Approximation)~\cite{Andersen} was applied. 
Experimentally obtained values of lattice constants for both Pr and Gd systems given in
Table~\ref{tab1} were used in our calculations. Atomic spheres radii were chosen as
$R$(Pr) = 3.91~a.u. and $R$(Fe) = 2.62~a.u. for Pr$_2$Fe$_{17}$ and $R$(Gd) = 3.72~a.u.
and $R$(Fe) = 2.66~a.u. for Gd$_2$Fe$_{17}$. Orbital basis consists of 6$s$, 6$p$, 5$d$,
and 4$f$ muffin-tin orbitals for Pr or Gd and 4$s$, 4$p$, and 3$d$ for Fe sites. The
calculations were performed with 32 irreducible {\bf k}-points (6$\times$6$\times$6
spacing) in the first Brillouin zone. Parameters of direct $U$ and exchange $J$ Coulomb
interactions for Gd and Pr ions were calculated by constrained LDA
method.~\cite{Gunnarsson01} For Gd$_2$Fe$_{17}$ we obtained $U_{Gd}$=6.7~eV and
$J_{Gd}$=0.7~eV (similar values were determined previously for elemental
Gd~\cite{Anisimov}), and for Pr$_2$Fe$_{17}$ $U_{Pr}$=4.9~eV and $J_{Pr}$=0.6~eV.
Corresponding values of $U$ and $J$ were applied to Gd and Pr compounds in frame of the
LSDA+U method. To note, in the present work we do not take into account local Coulomb
interaction on Fe ions since constrained LDA gives surprisingly large $U$ values.\cite{Gunnarsson91}
However, for elemental Fe it was shown that account of Coulomb
correlations is important to describe semiquantitatively temperature dependence of
magnetization.\cite{Lichtenstein01} But on the other hand such approach
reproduces main structures of LSDA DOS with slight modifications
of the Fe 3$d$ DOS and thus in our case will not affect resulting
dispersions of optical conductivity strongly.

Total magnetic moments per formula units obtained within LSDA+U method
(33.79~$\mu_B$ for Pr$_2$Fe$_{17}$ and 21.89~$\mu_B$ for Gd$_2$Fe$_{17}$)
are in good agreement with measured experimental data (see Table~\ref{tab1}).
Values of local magnetic moments for different sites for Pr$_2$Fe$_{17}$ and 
Gd$_2$Fe$_{17}$ compounds are listed in Table~\ref{tab2}.
Pr and Gd 4$f$-shells are computed to be fully polarised with local moments close to its ionic
values while Fe ions have local moments values almost equal to its elemental Fe magnitude.
It is remarkable that local moments on rare-earth sites during selfconsistent loops
become oppositely directed to those on Fe sites. One should also mention that initial
value of local moments on Pr and Gd ions was taken to be zero.
Thus R and Fe sublattices in our calculations are obtained to be antiferromagnetically ordered.

Calculated in the present work within the LSDA+U method partial Fe $4p, 3d$ and 
rare-earth $5d,4f$ DOS for spin-up ($\uparrow$) and spin-down
($\downarrow$) projections of local spin moments for Pr$_2$Fe$_{17}$ and
Gd$_2$Fe$_{17}$ are shown in Figs.~\ref{fig3}~and~\ref{fig4}. It is seen that main
spectral weight is located in the $E_F~\pm$~5~eV energy range around the Fermi level.
Structures of these DOS are rather similar for both compounds and are qualitatively close to
the DOS of ferromagnetic elemental iron in $bcc$ structure.~\cite{Gunnarsson02}
Both these spin-polarized
DOSs have a two-peak structure related to the $3d$ states of Fe ions. The Fermi level
$E_F$ set to zero lies near the minimum between these two peaks in the spin-down channel
and on the upper edge in the spin-up channel. ``Spin-up'' states of Fe ions are almost
completely occupied while ``spin-down'' are nearly half-filled. The narrow intensive peaks
at 3~eV ($\downarrow$) and 2.7~eV ($\uparrow$) (Pr$_2$Fe$_{17}$) also at $-7.3$~eV 
($\downarrow$) and 3.5~eV ($\uparrow$) (Gd$_2$Fe$_{17}$) belong to $4f$ states of
rare-earth ions. The intensities of Fe $4s, 4p$ and rare-earth $6s, 6p$, and $5d$
contributions to the DOS are considerably smaller.

\begin{table}
\caption{Calculated within the LSDA+U method
values of local magnetic moments for different sites in Pr$_2$Fe$_{17}$ and
Gd$_2$Fe$_{17}$ compounds.}
\begin{ruledtabular}
\begin{tabular}{ldld}
\multicolumn{2}{c}{Pr$_2$Fe$_{17}$}& \multicolumn{2}{c}{Gd$_2$Fe$_{17}$}\\
Site & \mbox{$M(\mu_B)$} & Site & \mbox{$M(\mu_B)$} \\
\hline
Pr($6c$) & -2.08 & Gd($2b$) &  -7.13 \\
Fe($6c$) & 2.21 & Gd($2d$) &  -7.20 \\
Fe($9d$) & 2.23 & Fe($4f$) &  2.31 \\
Fe($18f$) & 2.13 & Fe($6g$) &  2.10 \\
Fe($18h$) & 2.34 & Fe($12j$) &  2.40 \\
 & & Fe($12k$) & 1.87 \\
\end{tabular}
\end{ruledtabular}
\label{tab2}
\end{table}

\section{Analysis of optical conductivity structure}
\label{opt_theor}

\begin{figure}[]
\begin{center}
\epsfxsize=8.6cm
\epsfbox{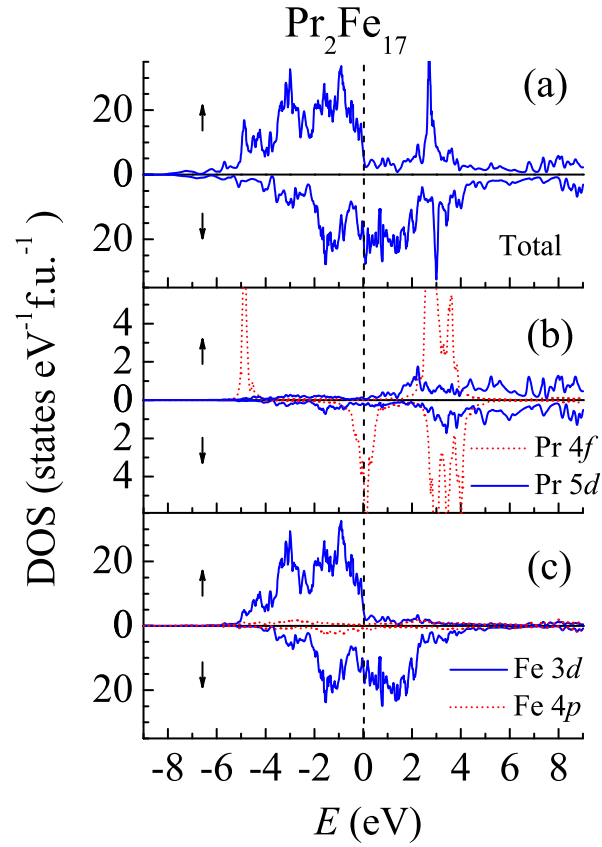}
\end{center}
\caption{(Colour online) Spin-resolved total (a) and partial $5d$ and $4f$ DOS of rare earth ion (Pr) (b) 
and $3d$ and $4p$ densities of states of Fe (c) for Pr$_2$Fe$_{17}$ compound obtained
from the LSDA+U calculation. DOSs were smoothed using adjacent averaging in the interval
0.1~eV. The Fermi level corresponds to zero.}
\label{fig3}
\end{figure}

\begin{figure}[]
\begin{center}
\epsfxsize=8.6cm
\epsfbox{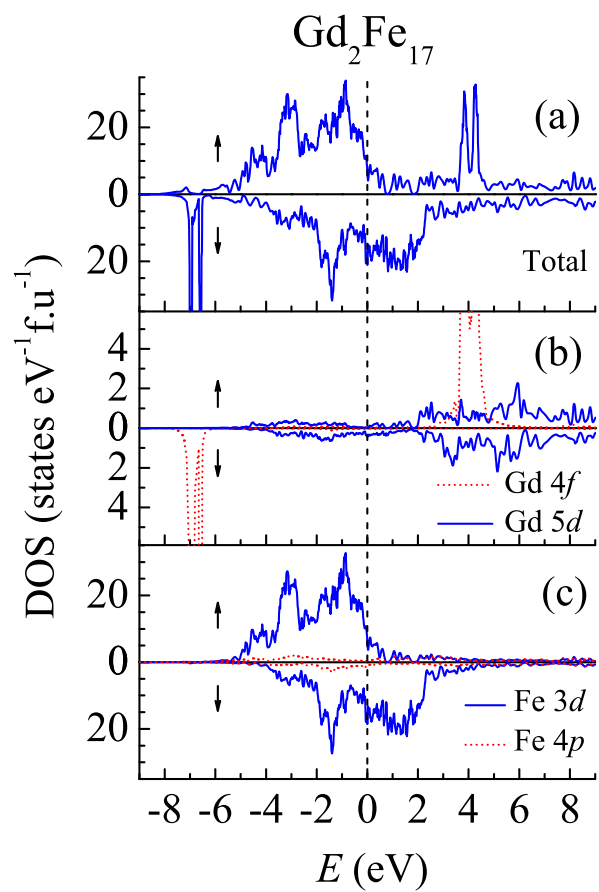}
\end{center}
\caption{(Colour online) The same as Fig.~\ref{fig3} but for Gd$_2$Fe$_{17}$.}
\label{fig4}
\end{figure}

Calculated LSDA+U band structures presented in the preceding section were used to interpret
experimental optical conductivity $\sigma(\omega)$ for both intermetallic systems under
consideration. In order to calculate theoretical optical conductivity
$\sigma_{theor}(\omega)$ and anatomize its ingredients we applied rather simplified
technique similar in spirit to Ref.~\cite{Spicer}. Namely, we computed following
convolutions representing all possible optical transitions
\begin{equation}
\sigma^{ll^{\prime }}_{is}(\omega)=-\frac 1{\hbar\omega}
\int_{E_F}^{E_F+\hbar\omega}N_{il}^s(E)N_{il^{\prime }}^s(E-\hbar\omega)dE,
\label{Occ}
\end{equation} 
where $N_{il}^s(E)$ and $N_{il^{\prime}}^s(E)$ are partial DOSs of the same ion $i$ with
the same spin $s$ and orbital quantum numbers $l$ and $l'$ shown in
Figs.~\ref{fig3}~and~\ref{fig4}. The orbital quantum numbers are related by dipol
selection rule $l-l'=\pm1$. Total theoretical optical conductivity is linear
combination of different contributions (\ref{Occ}):
\begin{equation}
\sigma_{theor}(\omega) = \sum_{\sigma, i, l-l'=\pm1}
             m_i s^{ll^{\prime }}_{is}(\omega),
\label{cond}
\end{equation} 
where $m_i$ is the degrees of site degeneracy (Wyckoff positions).

\begin{figure*}[]
\begin{center}
\epsfxsize=17.8cm
\epsfbox{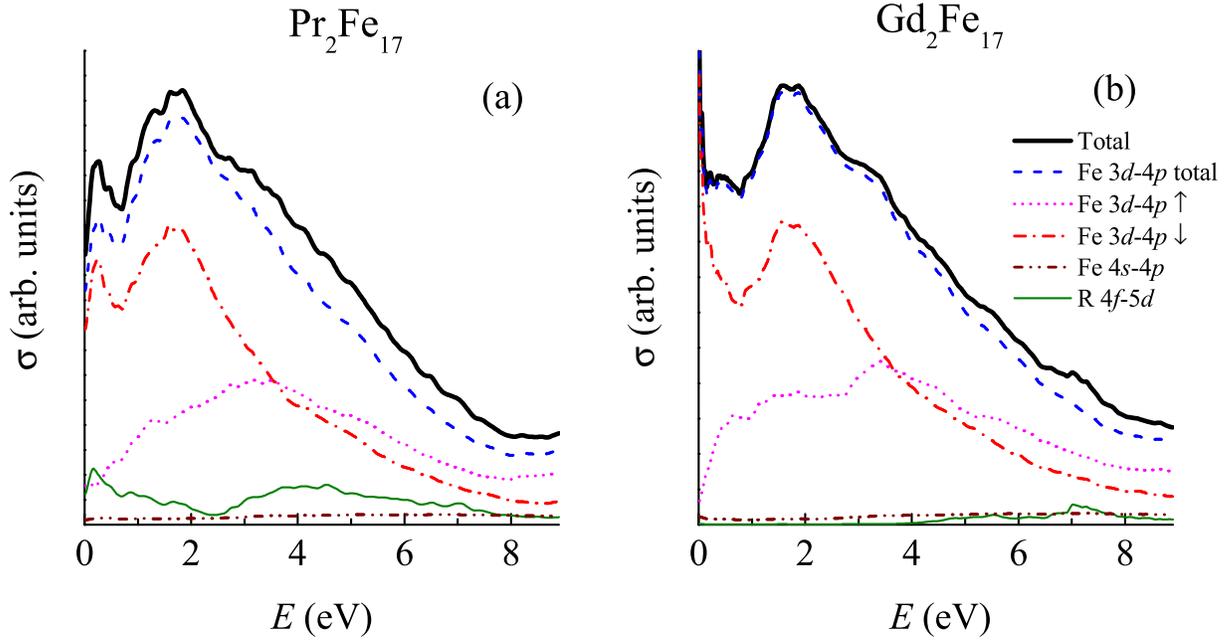}
\end{center}
\caption{(Colour online) Calculated total (thick solid lines) and partial contributions
in optical conductivity for Pr$_2$Fe$_{17}$ (a) and Gd$_2$Fe$_{17}$ (b) compounds.
Dashed, doted, and dash-doted lines show total, spin-up, and spin-down contributions from
$3d$-$4p$ transitions in Fe ions respectively. Dash-dot-doted lines represent contribution
from $4s$-$4p$ transitions in Fe, and thin solid lines correspond to $4f$-$5d$
transitions in rare-earth ions.}
\label{fig5}
\end{figure*}

Figs.~\ref{fig2} and~\ref{fig5} display corresponding $\sigma_{theor}(\omega)$ as
solid lines. The overall shape of optical conductivity dispersion curves for both
compounds exhibits a broad structure with the peak at $\sim$2~eV. On the whole the
theoretical calculations reproduce main features of the experimental $\sigma(\omega)$
curve (Fig.~\ref{fig2}): (i) the width of the intense absorption band with distinct
maximum, (ii) the sharp threshold in the 0.5--1~eV range and (iii) the gradual diminution
of the high-energy slope.

At the same time there are certain discrepancies. The theoretical peaks are slightly
shifted towards the higher energies in contrast to the experimental ones. Also
high-energy contributions at 4 eV are slightly overestimated. It may have following
reasons. As was shown by Lichtenstein $et~al.$\cite{Lichtenstein01} inclusion of local
dynamical Coulomb interactions into consideration in elemental Fe will lead to about 30\%
of correlation bandwidth narrowing. In our case DOS in the vicinity of the Fermi level
consists mostly from Fe 3$d$ states which are almost identical to those of elemental $bcc$
iron. Roughly one can say that for our intermetallic compounds it can slightly improve comparison
with experiment. First of all peak at 2 eV could be moved on 0.5--0.6 eV towards lower
energies because of correlation narrowing. Second spectral weight at 4 eV can be
transfered to lower energies and it can give high-energy tail of theoretical optical
conductivity closer to experimental one. Furthermore for both systems
$\sigma_{theor}(\omega)$ shows a significant interband absorption in the low-energy range
($\hbar\omega<$0.6~eV) which was not confirmed by the measurements. It is possible that
the low-energy absorption, predicted in theory, may be partially disguised in the
experimental $\sigma(\omega)$ curves by the strong Drude rising. 
The Drude contribution was not taken into consideration in our theoretical model.
 
Densities of states near the Fermi energy for both compounds mostly consist of the 3$d$-states of various crystallographically inequivalent iron ions (see Figs.~\ref{fig3} and \ref{fig4}). A line shape of our LSDA calculated Fe ions DOS is very similar to the known DOS of elemental Fe in the $bcc$ ($hcp$) phase.~\cite{Gunnarsson02,Wakon66} For the $bcc$ iron it was shown \cite{Bolotin69} that the optical conductivity determined using the Berglung-Spicer approach \cite{Spicer} agrees rather well with theoretical calculations accounting matrix elements\cite{Singh75} and also with experimental results.\cite{Shirokovskii82} This fact enables us to suppose that matrix elements formalism does not play an important role in case of intermetallic compounds R$_2$Fe$_{17}$.
In this work we can report our theoretical curves to be in a reasonable qualitative agreement with experimental data.

To analyze the line shape of experimental $\sigma(\omega)$ we provide detailed
description of different contributions to $\sigma_{theor}(\omega)$. As it is seen in
Fig.~\ref{fig5} for both systems biggest contribution to $\sigma_{theor}(\omega)$ comes
from $3d$--$4p$ interband optical transitions for Fe ions through the spin-down channel.
That gives a high absorption peak at $\sim$2~eV (see dotted curves in Fig.~\ref{fig5}).
These particular transitions mostly govern low-energy range behaviour below 1~eV as well.
Second largest contribution is $3d$--$4p$ transitions for Fe ions but in the spin-up
channel. In Fig.~\ref{fig5} one can see it as a rather broad structure with maxima at
$\sim$4~eV (dot-dashed curves). The contributions from Fe $4s$-$4p$ (Fig.~\ref{fig5},
dashed-dot-dot lines) and Pr(Gd) $4f$-$5d$ (Fig.~\ref{fig5}, thin solid line) transitions
are substantially smaller; and magnitude of convolutions of rare earths $5d$-$6p$ and
$6s$-$6p$ states is negligible.

\section{Summary}
\label{summary}

Comprehensive experimental investigation of structural, magnetic and optical properties
of the intermetallic compounds Pr$_2$Fe$_{17}$ and Gd$_2$Fe$_{17}$ was performed during
this work. Refined structural parameters are found to be in good agreement with previous
data. Measured magnetic properties: (i) Curie temperatures $T_C$=294K (466K) and (ii)
saturation magnetizations (at $T$=4.2K) 36.1 (21.2) $\mu_B$/f.u. for Pr-Fe (Gd-Fe) systems
respectively also agree well with available in the literature experimental data. We also
report measured for the first time optical constants $n$ and $k$ observed at room
temperature by ellipsometric Beattie technique in the spectral range of 0.22--15 $\mu$m.
This experimental data allows us to determine charge carriers parameters (plasma
$\Omega$ and relaxation  $\gamma$ frequencies) and optical conductivity $\sigma
(\omega)$.

To model magnetic and optical properties of Pr$_2$Fe$_{17}$ and Gd$_2$Fe$_{17}$ we did
self-consistent spin-resolved calculations within the LSDA+U method. Calculated by LSDA+U
method magnetic moments per formula unit describe well observed experimental values.
Furthermore experimental optical conductivity $\sigma (\omega)$ was interpreted in terms
of convolutions between partial densities of states for the same ion applying dipole
selection rule for orbital quantum number. Overall the line shape of experimental and
theoretical optical conductivity curves was found to be qualitatively very similar. By
anatomizing different contributions to $\sigma_{theor}(\omega)$ it was understood that
transitions between $3d$ and $4p$ states of Fe ions give the biggest contribution. The
intense peaks in optical conductivity below 1~eV and around 2~eV are predominantly formed
by the transitions in the ``spin-down'' channel while the second large contribution
from ``spin-up'' channel transitions provides broad feature with the maximum at about
4~eV. The other contributions from rare-earth $4f$-$5d$ and iron $4s$-$4p$ optical
transitions are almost negligible.

\section{Acknowledgments} This work was supported by Grants of Russian Foundation for
Basic Research 05-02-17244, 04-02-16086 and 05-02-16301 and in part by programs of the
Presidium of the Russian Academy of Sciences (RAS) ``Quantum macrophysics'' and of the
Division of Physical Sciences of the RAS ``Strongly correlated electrons in
semiconductors, metals, superconductors and magnetic materials''. Two of us (IN,AL)
acknowledge Dynasty Foundation and International Center for Fundamental Physics in
Moscow and Russian Science Support Foundation (IN).

\end {document}